\documentclass[conference]{IEEEtran}
\IEEEoverridecommandlockouts

\usepackage[english]{babel}

\usepackage{cite}
\usepackage{amsmath,amssymb,amsfonts}
\usepackage{algorithmic}
\usepackage{graphicx}
\usepackage{textcomp}
\usepackage{xcolor}
\usepackage{multirow}
\usepackage{caption}
\usepackage[caption=false,font=footnotesize]{subfig}
\usepackage[hidelinks]{hyperref}
\usepackage{tikz}
\usepackage{pgfgantt}
\usetikzlibrary{arrows.meta, positioning}

\tikzset{
    block/.style={rectangle, draw, rounded corners, align=center, minimum width=3.8cm, minimum height=1cm, font=\sffamily},
    arrow/.style={-{Latex[length=5mm, width=3mm]}, thick},
}

\def\BibTeX{{\rm B\kern-.05em{\sc i\kern-.025em b}\kern-.08em
    T\kern-.1667em\lower.7ex\hbox{E}\kern-.125emX}}
\begin{document}

\title{ XAI-Driven Deep Learning for Protein Sequence Functional Group Classification
}

\author{
\IEEEauthorblockN{Pratik Chakraborty}
\IEEEauthorblockA{
\textit{Manipal Institute of Technology, Manipal, India} \\
Email: pratikchakra18@gmail.com}
\and
\IEEEauthorblockN{Aryan Bhargava}
\IEEEauthorblockA{
\textit{Manipal Institute of Technology, Manipal, India} \\
Email: aryanbhargava85@gmail.com}
}

\maketitle

\begin{abstract}
Proteins perform essential biological functions, and accurate classification of their sequences is critical for understanding structure–function relationships, enzyme mechanisms, and molecular interactions. This study presents a deep learning-based framework for functional group classification of protein sequences derived from the Protein Data Bank (PDB). Four architectures were implemented: Convolutional Neural Network (CNN), Bidirectional Long Short-Term Memory (BiLSTM), CNN-BiLSTM hybrid, and CNN with Attention. Each model was trained using k-mer integer encoding to capture both local and long-range dependencies. Among these, the CNN achieved the highest validation accuracy of 91.8\%, demonstrating the effectiveness of localized motif detection. Explainable AI techniques, including Grad-CAM and Integrated Gradients, were applied to interpret model predictions and identify biologically meaningful sequence motifs. The discovered motifs, enriched in histidine, aspartate, glutamate, and lysine, represent amino acid residues commonly found in catalytic and metal-binding regions of transferase enzymes. These findings highlight that deep learning models can uncover functionally relevant biochemical signatures, bridging the gap between predictive accuracy and biological interpretability in protein sequence analysis.
\end{abstract}


\section{Introduction}
\label{sec:introduction}

Proteins are fundamental biological macromolecules that play essential roles in virtually all cellular processes, including catalyzing biochemical reactions, transmitting signals, transporting molecules, and maintaining structural integrity. Understanding the relationship between a protein’s amino acid
sequence and its biochemical function is therefore a central challenge in computational biology. Accurate functional classification of protein sequences not only aids in annotating novel proteins discovered through large-scale genomic studies but also supports downstream applications in drug discovery,enzyme engineering, biomarker identification, and disease
diagnostics.

Early computational studies on protein sequence analysis laid the statistical and algorithmic foundations for modern bioinformatics. Classical works introduced quantitative approaches for analyzing residue composition, sequence biases, and non-random motif occurrences within amino acid chains~\cite{brendel1992methods}. These statistical and Markov-based frameworks helped identify conserved and functionally significant regions, enabling the first algorithmic strategies for motif discovery, sequence alignment, and family classification. Building upon this foundation, subsequent
research established the concept of protein sequence motifs as short, conserved residue patterns linked to specific structural or functional roles~\cite{bork1996protein}. Such motifs, which often
corresponding to enzyme active sites, metal-binding residues, or transmembrane regions, serve as diagnostic signatures of protein families and functional domains. Later studies expanded
motif analysis to include membrane-associated proteins, where recurrent sequence and structural motifs mediate helix--helix interactions and signaling processes~\cite{moore2008protein}.

With the advent of high-throughput sequencing and the
accumulation of vast protein databases such as UniProt and the Protein Data Bank (PDB), machine learning (ML) and deep learning (DL) methods have become indispensable for sequence-based protein classification. Early ML approaches
relied on handcrafted descriptors such as amino acid
composition, pseudo-amino acid composition, and
position-specific scoring matrices. However, recent surveys demonstrate that deep learning architectures can learn high-dimensional representations directly from raw amino acid sequences, eliminating the need for manual feature engineering~\cite{ao2022biological},\cite{dhanuka2023comprehensive}. Convolutional neural networks (CNNs) have proven effective for capturing local motif-like dependencies, while recurrent models such as long short-term memory (LSTM) networks capture sequential context and long-range dependencies across
residues~\cite{wen2020deep,cui2021sequence}. These models have achieved substantial improvements in classification accuracy and generalization over traditional methods.

Inspired by the success of natural language processing
(NLP), recent advances treat protein sequences as a ``language
of life,'' where amino acids act as tokens and sequence context
defines semantic meaning~\cite{ofer2021language}. Transformer architectures and protein language models have emerged as a new paradigm, learning contextual embeddings through
self-supervised pretraining on millions of sequences. These models, such as ProteinBERT, ProtBERT, ProtT5, and ESM-series, capture both local and global relationships within protein sequences and have demonstrated remarkable success in functional prediction, structure estimation, and protein
design~\cite{hu2024advances,jisna2021protein}. Surveys show that deep and transformer-based models are increasingly used for diverse tasks spanning sequence modeling, structure prediction, and multi-modal protein representation learning.

Despite these advances, interpretability remains a critical challenge. Deep learning models often behave as black boxes, offering limited insight into which residues or subsequences drive their predictions. Explainable AI (XAI) methods address this gap by providing residue-level attributions that reveal how
models capture functional or structural cues. Integrated Gradients and Gradient-weighted Class Activation Mapping
(Grad-CAM) have been successfully adapted to protein
sequence models, allowing visualization of motif-level
importance and biologically relevant signal propagation within sequences. Recent work emphasizes that biologically plausible interpretations—those aligning with known motifs or catalytic
sites—are vital for building trust in predictive models and for guiding protein engineering decisions.

While deep learning and XAI have been applied to protein function prediction, comprehensive comparative studies that evaluate multiple model architectures alongside interpretability remain limited. Furthermore, many prior works emphasize
predictive accuracy without systematically analyzing how different architectures focus on distinct biochemical patterns.
To address these gaps, our study makes the following
contributions:
\begin{itemize}
    \item We conduct a comprehensive comparative analysis of deep learning architectures like CNN, BiLSTM, CNN-BiLSTM, and CNN-Attention for protein functional group classification using PDB-derived sequences.
    \item We evaluate how each model captures biologically meaningful motifs and long-range dependencies through explainability techniques including Grad-CAM and Integrated Gradients.
    \item We identify consensus motifs across models that correspond to conserved catalytic or binding residues, providing insights into the biological plausibility of learned
    representations.
\end{itemize}

The remainder of this paper is organized as follows:
Section~\ref{sec:related_work} reviews related work on deep learning, transformer models, and XAI techniques for protein sequence classification. Section~\ref{sec:methodology}
describes the dataset, preprocessing, and model architectures.
Section~\ref{sec:results} presents the experimental results and
interpretability analyses. Finally, Section~\ref{sec:conclusion}
concludes the paper and discusses future research directions.

\section{Related Work}
\label{sec:related_work}

Protein sequence classification has been widely explored using deep learning approaches, ranging from CNNs and RNNs to more advanced transformer-based protein language models. Recently, explainable AI (XAI) techniques have been introduced to improve interpretability, enabling models to highlight biologically meaningful motifs and functional patterns. In this section, we review prior work under three themes: deep learning methods, transformer and protein language models, and XAI techniques for protein sequence functional group classification.

\subsection{Deep Learning Methods}

Deep learning has become a central tool for protein sequence classification, enabling models to learn informative representations directly from amino acid sequences rather than relying entirely on handcrafted features. Recurrent neural networks, particularly LSTM-based approaches, have been employed to capture sequential dependencies and achieve effective function prediction \cite{liu2017deep,sekhar2021protein}. Convolutional neural networks have also shown strong performance by identifying local sequence motifs and, in many cases, surpassing traditional machine learning baselines in protein family classification \cite{zhang2020protein,wang2022deep}. More recent studies have investigated hybrid strategies that combine encoding schemes such as k-mers, count vectorization, and substitution matrices with deep models, underlining the importance of representation choice in achieving high classification accuracy \cite{tasnim2024protein,perveen2025protein}. Large-scale experiments further demonstrated that deep neural networks can generalize across vast protein datasets, successfully annotating families and uncovering functional patterns beyond what alignment-based methods could capture \cite{bileschi2022using}. These developments have firmly established deep learning as a foundation for protein sequence analysis and set the stage for transformer-based architectures and explainable AI techniques.  

\subsection{Transformers and Protein Language Models}

Transformer architectures and self-supervised protein language models have significantly advanced sequence-based protein classification by learning contextual embeddings from large-scale unlabeled protein datasets. ProteinBERT is one such model, inspired by BERT, that combines masked language modeling with Gene Ontology prediction during pretraining and incorporates both global and local representations for efficient handling of long sequences \cite{brandes2022proteinbert}. ProtTrans introduced a family of models including ProtBERT and ProtT5, pretrained on millions of sequences from UniProt, which demonstrated that embeddings derived from transformers can be effectively fine-tuned for a wide range of downstream classification and prediction tasks \cite{elnaggar2021prottrans}.  

Subsequent work has aimed to extend protein language models by integrating structural information and optimizing attention mechanisms. ProSST incorporates quantized structure tokens and disentangled attention, explicitly modeling the relationship between sequence and structural representations to improve functional and stability-related predictions \cite{li2024prosst}. UDSMProt demonstrated the potential of universal deep sequence models, showing that transformer-based embeddings could generalize across diverse protein classification benchmarks without heavy task-specific tuning \cite{strodthoff2020udsmprot}. Newer models such as Ankh have focused on architectural optimizations, enabling more efficient and general-purpose protein sequence modeling \cite{elnaggar2023ankh}.  

Beyond classification, transformer language models have also been applied to protein design and large-scale structure prediction. ProtGPT2 was trained as an autoregressive model for unsupervised protein design, generating novel sequences with properties resembling natural proteins \cite{ferruz2022protgpt2}. Evolutionary-scale modeling with ESM-based transformers has even achieved atomic-level structure prediction, showing that sequence-only models can capture deep biological constraints traditionally requiring structural input \cite{lin2023evolutionary}. Together, these advances highlight the transformative impact of protein language models on functional prediction, while also raising new challenges regarding the interpretability of their learned representations.  

\subsection{XAI Techniques}

The complexity of deep learning and transformer-based models for protein sequence classification has led to growing interest in explainable AI methods that make predictions more interpretable. Attribution approaches such as Integrated Gradients \cite{sundararajan2017axiomatic} and gradient-weighted class activation mapping (Grad-CAM) \cite{selvaraju2017grad} have been adapted to highlight residues and sequence regions that contribute most to classification outcomes. Attention-based methods have also been examined, where quantifying the flow of attention provides insight into how contextual dependencies across long sequences are modeled \cite{abnar2020quantifying}.  

Building on these foundations, transformer models for protein function prediction have been analyzed using Integrated Gradients to identify residue-level importance and specialized attention heads \cite{wenzel2024insights}. Comparative evaluations of multiple attribution techniques have shown how different methods capture distinct biochemical properties and functional signals \cite{fazel2025explainability}. In protein engineering, workflows have been proposed that combine predictive models with post hoc explanations to guide rational design decisions \cite{medina2025interpretable}. Multi-stage attention mechanisms that integrate both sequence and structural information have further demonstrated how attention maps can highlight the most influential features for functional prediction \cite{liu2025multi}.  

Survey work has provided a broader perspective, outlining categories of explainability methods used in bioinformatics, identifying evaluation challenges, and emphasizing the importance of biological plausibility when interpreting model outputs \cite{budhkar2025demystifying}. Collectively, these contributions illustrate how attribution, attention analysis, and interpretability-oriented model designs are becoming central to the application of deep learning in protein sequence functional group classification.

\section{Methodology}
\label{sec:methodology}

\begin{figure}[ht]
\centering
\resizebox{0.49\textwidth}{!}{%
\begin{tikzpicture}[node distance=1.8cm and 2.8cm]

\node[block] (dataset) {PDB Dataset};
\node[block, below=of dataset] (filter) {Filtering \\ Top 10 Functional Classes};
\node[block, below=of filter] (kmers) {Generate k-mers};
\node[block, below=of kmers] (intencode) {Integer Encoding};

\node[block, below=3cm of intencode, xshift=-6cm] (cnn) {CNN};
\node[block, below=3cm of intencode, xshift=-2cm] (bilstm) {BiLSTM};
\node[block, below=3cm of intencode, xshift=2cm] (cnnbilstm) {CNN-BiLSTM};
\node[block, below=3cm of intencode, xshift=6cm] (cnnattn) {CNN-Attention};
\node[block, below=12cm of kmers, align=center] (output) {Protein Functional Group Classification \\ (10 Classes)};

\draw[arrow] (dataset) -- (filter);
\draw[arrow] (filter) -- (kmers);
\draw[arrow] (kmers) -- (intencode);

\draw[arrow] (intencode) -- (cnn.north);
\draw[arrow] (intencode) -- (bilstm.north);
\draw[arrow] (intencode) -- (cnnbilstm.north);
\draw[arrow] (intencode) -- (cnnattn.north);

\draw[arrow] (cnn) -- (output.north);
\draw[arrow] (bilstm) -- (output.north);
\draw[arrow] (cnnbilstm) -- (output.north);
\draw[arrow] (cnnattn) -- (output.north);

\end{tikzpicture}%
}
\caption{Proposed pipeline for protein functional group classification using four deep learning architectures.}
\label{fig:methodology}
\end{figure}
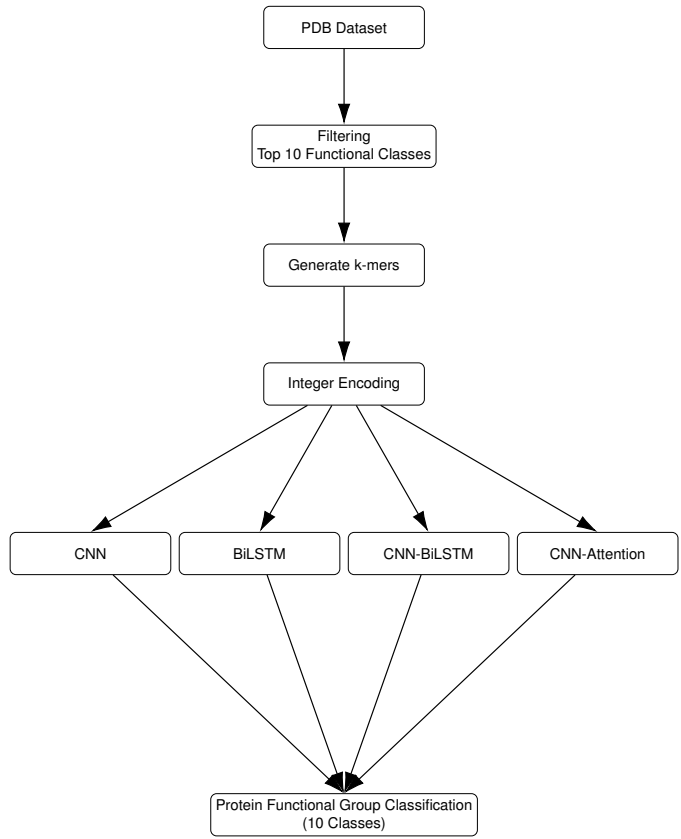

This section presents the methodology employed for protein sequence functional group classification. The objective is to categorize protein sequences from the Protein Data Bank (PDB) dataset into ten of the most frequent functional classes, such as hydrolases, transferases, oxidoreductases, and lyases \cite{berman2000protein}. The proposed framework combines sequence preprocessing using k-mer encoding, multiple deep learning architectures for classification, and explainable AI (XAI) techniques for biological interpretability. The overall workflow is illustrated in Figure~\ref{fig:methodology}.

\subsection{Dataset}

This study makes use of the Structural Protein Sequences dataset available on Kaggle, which is derived from the Research Collaboratory for Structural Bioinformatics (RCSB) Protein Data Bank (PDB). The PDB is a comprehensive, publicly accessible archive of experimentally determined three-dimensional structural data for biological macromolecules, including proteins, DNA, and RNA. Structural information is obtained using techniques such as X-ray crystallography, nuclear magnetic resonance (NMR) spectroscopy, and cryo-electron microscopy.  

The dataset contains more than 400,000 entries and is organized into two primary components. The first is a metadata file, which provides high-level structural and experimental details for each entry, including protein classification (for example, hydrolase, transferase), macromolecule type, structure resolution, crystallization method, and publication year. The second is a sequence file, which includes raw biological sequences for proteins, DNA, RNA, and hybrid macromolecules, along with chain identifiers and residue counts.  

For the purposes of this work, the analysis is restricted to entries where the macromolecule type is Protein. To ensure balanced evaluation and biologically meaningful categorization, the study focuses on the ten most frequent functional classes in the dataset. Table~\ref{tab:top10classes} summarizes these categories and their frequencies.  

\begin{table}[h]
\centering
\caption{Top 10 protein functional classes used in this study.}
\label{tab:top10classes}
\begin{tabular}{|l|c|}
\hline
Class                          & Frequency \\
\hline
Hydrolase                      & 46336 \\
Transferase                    & 36424 \\
Oxidoreductase                 & 34321 \\
Immune System                  & 15615 \\
Lyase                          & 11682 \\
Hydrolase/Hydrolase Inhibitor  & 11218 \\
Transcription                  & 8919  \\
Viral Protein                  & 8495  \\
Transport Protein              & 8371  \\
Virus                          & 6972  \\
\hline
\end{tabular}
\end{table}

\subsection{Preprocessing}

The preprocessing pipeline involved two key stages: sequence segmentation and integer encoding. Each protein sequence was segmented into overlapping k-mers of length five (\(k=5\)), capturing local residue-level dependencies. Each unique k-mer was then assigned an integer index based on frequency ranking, with rare k-mers truncated to maintain a fixed vocabulary size.  

These integer-encoded sequences were padded or truncated to a maximum sequence length of 300 to ensure uniform input dimensions across models. This representation allows the embedding layers of neural networks to learn distributed vector representations for k-mers, effectively capturing sequence order and contextual relationships.

\subsection{Models Used}

Four baseline deep learning architectures were implemented to model protein sequences and predict functional classes. Each model captures different aspects of sequence information, from local motif extraction to long-range dependencies.

\begin{itemize}
    \item \textbf{Convolutional Neural Network (CNN):} Applies one-dimensional convolutional filters to identify local motif patterns indicative of functional and structural characteristics.  
    \item \textbf{Bidirectional Long Short-Term Memory (BiLSTM):} Processes sequences in both forward and backward directions, enabling the model to capture long-range dependencies between amino acids.  
    \item \textbf{CNN-BiLSTM:} Integrates convolutional layers for local motif detection with BiLSTM layers for sequential modeling, combining short-term and long-term contextual learning.  
    \item \textbf{CNN-Attention:} Enhances convolutional feature extraction with an attention mechanism, allowing the network to assign greater importance to functionally relevant subsequences.  
\end{itemize}

The parameter counts for each model are summarized in Table~\ref{tab:model_params}, reflecting the computational complexity and capacity of each architecture.

\begin{table}[h]
\centering
\caption{Trainable parameters of each deep learning model.}
\label{tab:model_params}
\begin{tabular}{|l|c|}
\hline
\textbf{Model} & \textbf{Trainable Parameters (Millions)} \\
\hline
CNN & 6.65 M \\
BiLSTM & 8.77 M \\
CNN-BiLSTM & 7.28 M \\
CNN-Attention & 6.57 M \\
\hline
\end{tabular}
\end{table}

This comparison highlights that the BiLSTM model possesses the highest parameter count, reflecting its increased capacity to model long-range dependencies, whereas the CNN-Attention model achieves efficient representation learning with relatively fewer parameters.

\subsection{Training and Hyperparameter Tuning}

All models were trained under a unified experimental setup to ensure a fair comparison across architectures. Each model was optimized using the Adam optimizer with carefully tuned hyperparameters to balance convergence speed and generalization. Early stopping was applied to prevent overfitting, monitoring validation performance over epochs. The training configuration and hyperparameters are summarized in Table~\ref{tab:hyperparams}.  

\begin{table}[h]
\centering
\caption{Training configuration and hyperparameters used for all models.}
\label{tab:hyperparams}
\begin{tabular}{|l|c|}
\hline
\textbf{Parameter} & \textbf{Value} \\
\hline
Optimizer & Adam \\
Batch Size & 64 \\
Epochs & 30  \\
Early Stopping Patience & 5 epochs \\
Learning Rate & $1 \times 10^{-3}$ \\
Dropout Rate & 0.3 \\
\hline
\end{tabular}
\end{table}

This consistent setup allowed all the deep learning architectures to be evaluated under comparable conditions while ensuring stability and reproducibility of results.

\subsection{Explainable AI Techniques}

To provide biological interpretability, two complementary XAI techniques were employed to analyze model predictions and highlight functionally relevant subsequences:

\begin{itemize}
    \item \textbf{Integrated Gradients (IG):} Quantifies the contribution of each input k-mer to the model’s prediction by integrating gradients along a path from a baseline input to the actual input. This yields fine-grained importance scores, revealing critical residues or motifs influencing classification.  
    \item \textbf{Gradient-weighted Class Activation Mapping (Grad-CAM):} Adapted for one-dimensional convolutional layers, Grad-CAM identifies salient regions in the sequence that contribute most to the predicted functional class. This provides a coarse-grained, region-level interpretability complementary to IG.  
\end{itemize}

The combined application of IG and Grad-CAM offers both residue-level and region-level interpretability, enabling the identification of key functional motifs learned by each deep learning architecture.

\subsection{Evaluation Metrics}

Model performance was evaluated using multiple quantitative and qualitative metrics to assess both predictive accuracy and interpretability. Accuracy, precision, recall, and F1-score were computed to provide a balanced evaluation of overall and class-wise performance. Additionally, confusion matrices were generated for each model to visualize class-level prediction patterns and identify sources of misclassification.

To assess the stability and convergence behavior of each architecture, training and validation accuracy curves were plotted across epochs, providing insights into learning dynamics and generalization trends.

Beyond standard performance metrics, explainable AI (XAI) evaluations were conducted using Integrated Gradients (IG) and Gradient-weighted Class Activation Mapping (Grad-CAM). These attribution techniques enabled qualitative assessment of model interpretability by highlighting key amino acid motifs and subsequences contributing to classification decisions.

Together, these quantitative metrics and interpretability analyses provided a comprehensive evaluation framework, ensuring that each model was assessed not only in terms of predictive performance but also in terms of biological explainability and reliability.
 
\section{Results and Discussion}
\label{sec:results}
This section presents the results obtained from the experiments conducted on the PDB dataset using the four deep learning architectures and explainable AI techniques. The section begins with an analysis of the predictive performance of the models, followed by interpretability insights derived from Integrated Gradients and Grad-CAM visualizations, which highlight biologically meaningful motifs within protein sequences.

\subsection{Performance of the Deep Learning Models}

Table~\ref{tab:val_acc} summarizes the validation accuracies achieved by all four models on the protein functional group classification task. Among them, the CNN model achieved the highest validation accuracy of 91.80\%, indicating its strong ability to extract local motifs that are functionally discriminative. The BiLSTM model followed closely, achieving comparable performance, while hybrid and attention-based models exhibited slightly lower accuracies, likely due to overfitting or added architectural complexity.

\begin{table}[h]
\centering
\caption{Validation accuracy of each deep learning model on the PDB dataset.}
\label{tab:val_acc}
\begin{tabular}{|l|c|}
\hline
\textbf{Model} & \textbf{Validation Accuracy (\%)} \\
\hline
CNN & \textbf{91.80} \\
BiLSTM & 91.75 \\
CNN-BiLSTM & 91.30 \\
CNN-Attention & 91.17 \\
\hline
\end{tabular}
\end{table}

\begin{figure}[h]
\centering
\subfloat[CNN]{%
    \includegraphics[width=0.23\textwidth]{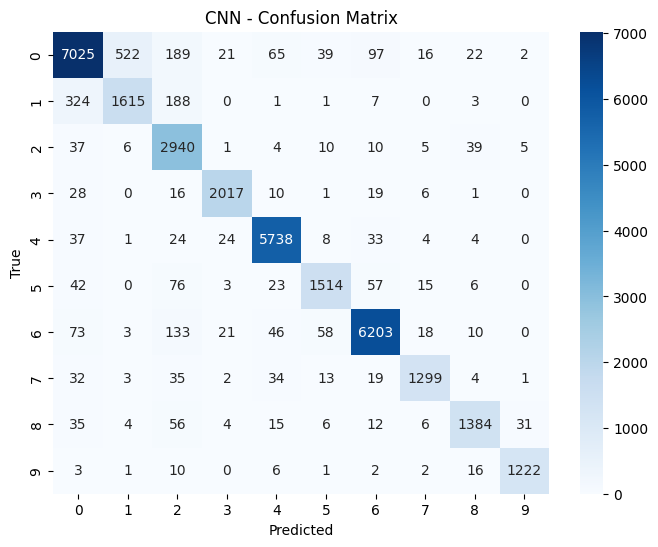}
}
\hfill
\subfloat[BiLSTM]{%
    \includegraphics[width=0.23\textwidth]{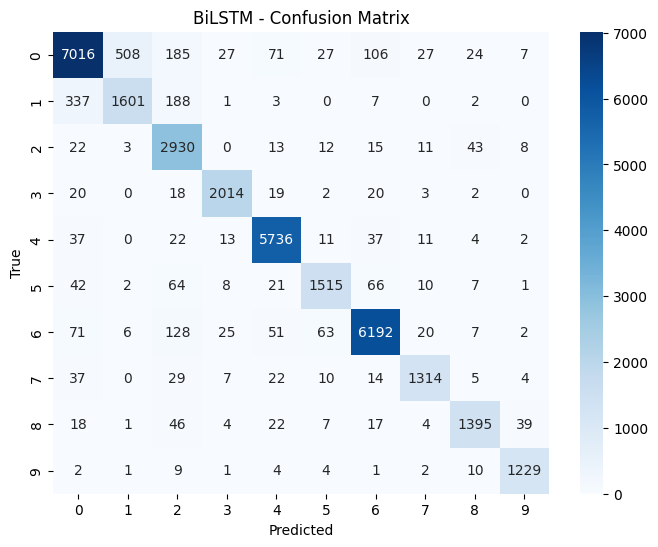}
}\\
\subfloat[CNN-BiLSTM]{%
    \includegraphics[width=0.23\textwidth]{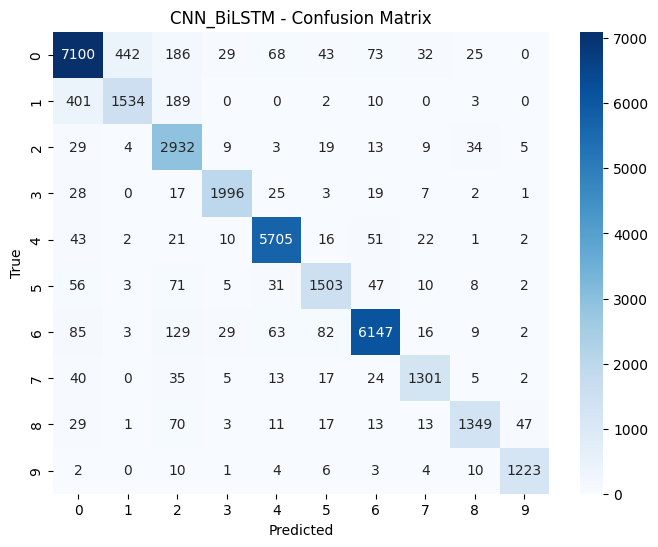}
}
\hfill
\subfloat[CNN-Attention]{%
    \includegraphics[width=0.23\textwidth]{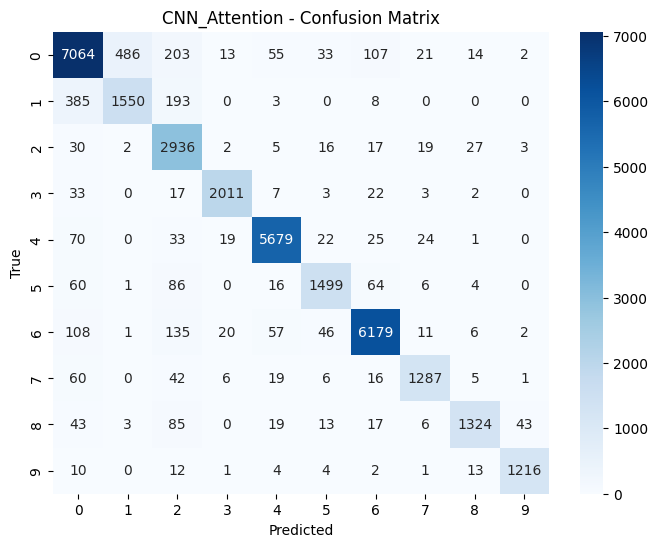}
}
\caption{Confusion matrices for each deep learning model on the test set.}
\label{fig:conf_matrices}
\end{figure}

Figure~\ref{fig:conf_matrices} presents the confusion matrices for all models, illustrating class-wise prediction performance and common misclassification patterns. These matrices reveal that the models generally perform well across all functional categories, with occasional overlaps between classes such as transferases and hydrolases, which are known to share structural similarities.

\begin{figure}[h]
\centering
\subfloat[CNN]{%
    \includegraphics[width=0.23\textwidth]{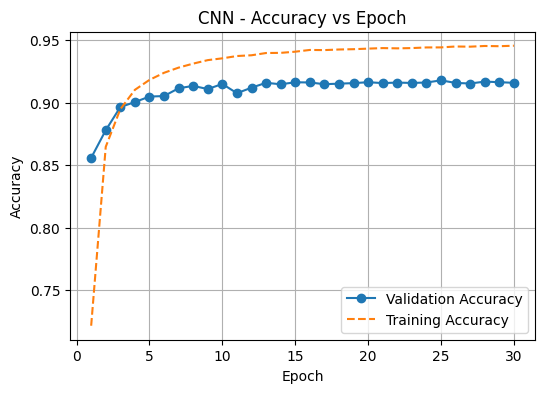}
}
\hfill
\subfloat[BiLSTM]{%
    \includegraphics[width=0.23\textwidth]{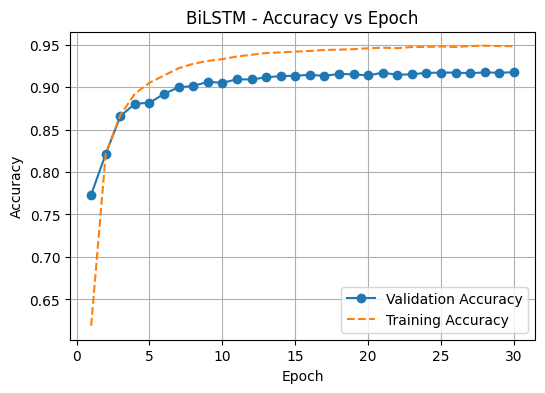}
}\\
\subfloat[CNN-BiLSTM]{%
    \includegraphics[width=0.23\textwidth]{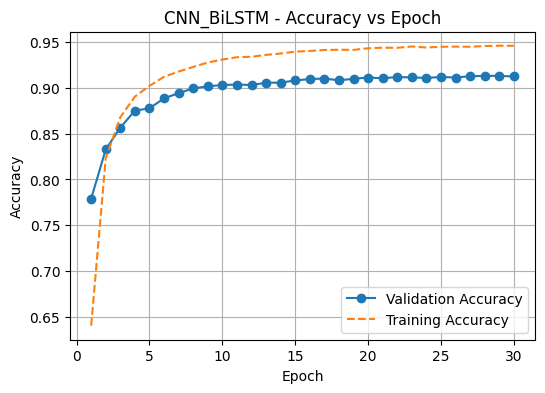}
}
\hfill
\subfloat[CNN-Attention]{%
    \includegraphics[width=0.23\textwidth]{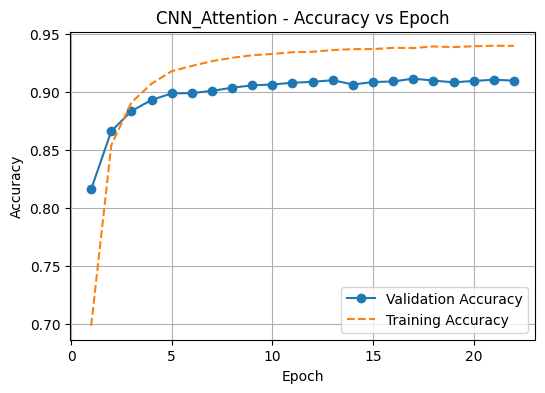}
}
\caption{Training and validation accuracy vs. epoch for each model.}
\label{fig:accuracy_curves}
\end{figure}

To further examine the learning behavior of each model, Figure~\ref{fig:accuracy_curves} shows the training and validation accuracy curves across epochs. All models demonstrate smooth convergence with minimal overfitting, validating the effectiveness of the chosen hyperparameters and early stopping strategy. Across architectures, the curves exhibit consistent learning dynamics characterized by a sharp increase in accuracy during the first few epochs, followed by a gradual stabilization phase after approximately ten epochs as the models approach convergence. The validation accuracy closely tracks the training curve, maintaining a small generalization gap of about 2–3\%, indicating strong regularization and robust generalization to unseen data.

\begin{figure*}[t]
\centering
\subfloat[Grad-CAM — CNN]{\includegraphics[width=0.47\textwidth]{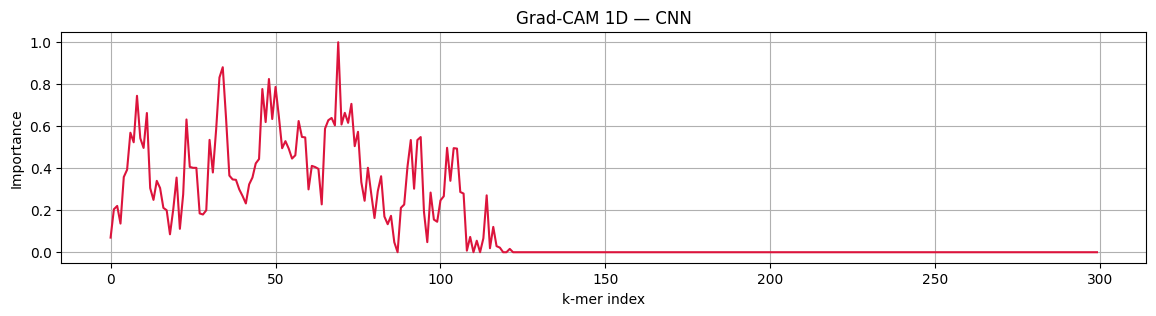}}
\hfill
\subfloat[Grad-CAM — CNN-BiLSTM]{\includegraphics[width=0.47\textwidth]{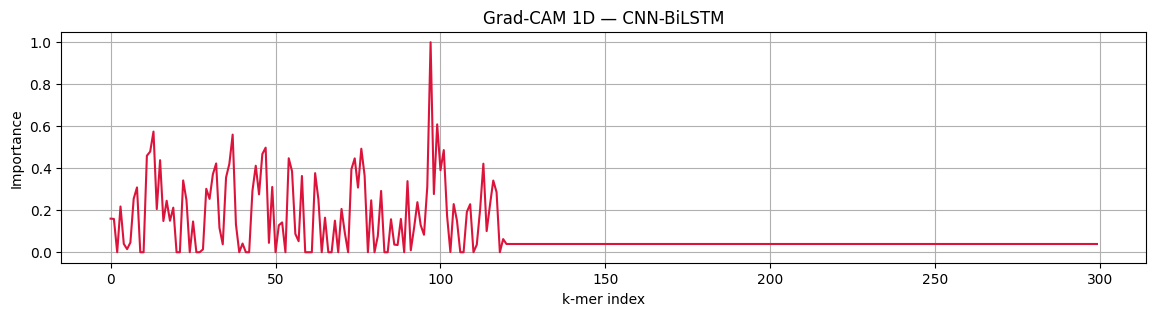}}\\
\subfloat[Integrated Gradients — BiLSTM]{\includegraphics[width=0.47\textwidth]{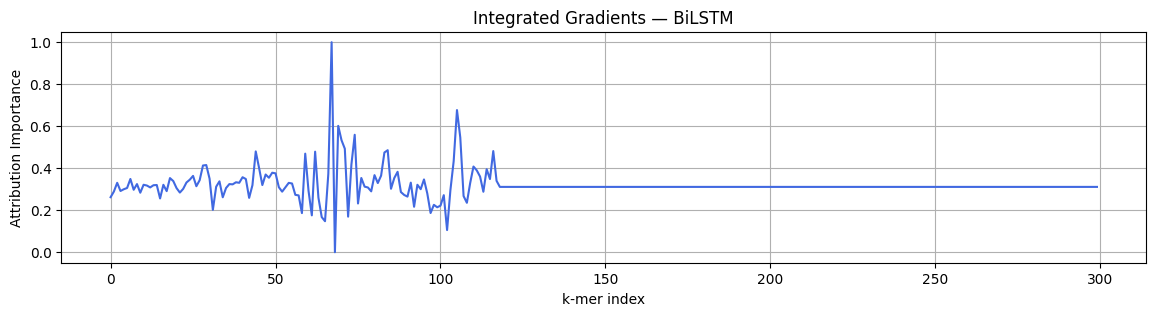}}
\hfill
\subfloat[Integrated Gradients — CNN-Attention]{\includegraphics[width=0.47\textwidth]{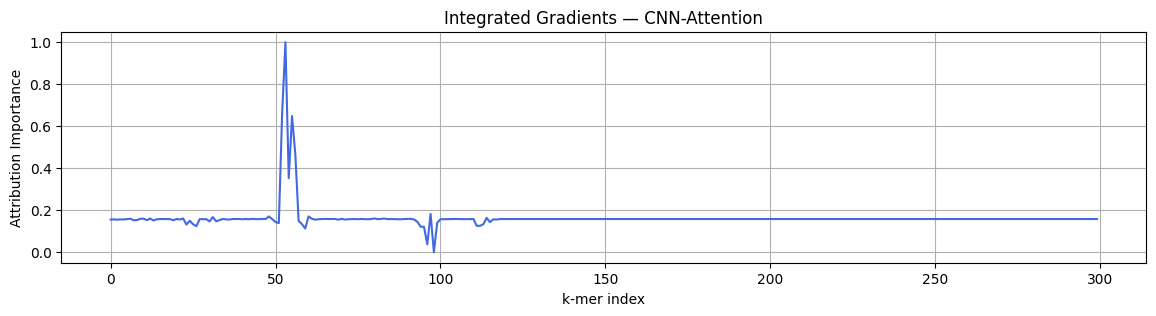}}
\caption{Explainability visualizations for the Transferase class using Grad-CAM and Integrated Gradients across four deep learning architectures.}
\label{fig:xai_curves}
\end{figure*}

\begin{table*}[t]
\centering
\caption{Top 10 motifs identified for the Transferase class by each model with their XAI technique, sequence position, and normalized importance score.}
\label{tab:xai_motifs_full}
\resizebox{0.37\textwidth}{!}{
\begin{tabular}{|l|l|l|c|c|}
\hline
\textbf{Model} & \textbf{XAI Technique} & \textbf{Motif} & \textbf{Position} & \textbf{Score} \\ \hline

\multirow{10}{*}{CNN} & \multirow{10}{*}{Grad-CAM} 
& HQLLH & 69--74 & 1.0000 \\
& & KVYNG & 34--39 & 0.8810 \\
& & VKVYN & 33--38 & 0.8321 \\
& & DRLYA & 48--53 & 0.8249 \\
& & LYASA & 50--55 & 0.7871 \\
& & HIDRL & 46--51 & 0.7771 \\
& & IVKDE & 8--13 & 0.7449 \\
& & HELVE & 73--78 & 0.7065 \\
& & LLHEL & 71--76 & 0.6637 \\
& & DEEVK & 11--16 & 0.6629 \\ \hline

\multirow{10}{*}{BiLSTM} & \multirow{10}{*}{Integrated Gradients} 
& KFHQL & 67--72 & 1.0000 \\
& & TVKPV & 105--110 & 0.6769 \\
& & HQLLH & 69--74 & 0.6017 \\
& & ELVEK & 74--79 & 0.5593 \\
& & VKPVI & 106--111 & 0.5467 \\
& & QLLHE & 70--75 & 0.5335 \\
& & LLHEL & 71--76 & 0.4929 \\
& & GHIYF & 84--89 & 0.4858 \\
& & ENPRP & 116--121 & 0.4816 \\
& & NEHID & 44--49 & 0.4800 \\ \hline

\multirow{10}{*}{CNN-BiLSTM} & \multirow{10}{*}{Grad-CAM} 
& RAHQF & 97--102 & 1.0000 \\
& & HQFPE & 99--104 & 0.6089 \\
& & EVKID & 13--18 & 0.5745 \\
& & NGEMF & 37--42 & 0.5601 \\
& & IDRLY & 47--52 & 0.4979 \\
& & VEKNE & 76--81 & 0.4929 \\
& & FPENT & 101--106 & 0.4864 \\
& & EEVKI & 12--17 & 0.4792 \\
& & HIDRL & 46--51 & 0.4675 \\
& & DEEVK & 11--16 & 0.4595 \\ \hline

\multirow{10}{*}{CNN-Attention} & \multirow{10}{*}{Integrated Gradients} 
& SAEKI & 53--58 & 1.0000 \\
& & ASAEK & 52--57 & 0.6604 \\
& & EKIRI & 55--60 & 0.6485 \\
& & KIRIT & 56--61 & 0.4650 \\
& & AEKIR & 54--59 & 0.3528 \\
& & RAHQF & 97--102 & 0.1823 \\
& & TIPYT & 60--65 & 0.1704 \\
& & DRLYA & 48--53 & 0.1703 \\
& & EVVKV & 31--36 & 0.1674 \\
& & TKENP & 114--119 & 0.1637 \\ \hline

\end{tabular}
}
\end{table*}

Overall, the results demonstrate that convolution-based models are particularly effective in recognizing local sequence patterns that correlate with protein functionality, while recurrent and hybrid approaches provide comparable yet more computationally intensive alternatives.

\subsection{Explainable AI Analysis of Transferase Motifs}
\label{sec:xai_transferase}

To interpret the learned representations and validate biological relevance, explainable AI (XAI) techniques were applied to the Tranferase class using Grad-CAM for convolutional architectures and Integrated Gradients for recurrent and attention-based models. Figure~\ref{fig:xai_curves} illustrates the one-dimensional attribution profiles, where sharp spikes denote $k$-mers with high class-discriminative importance.

The motifs highlighted across all four architectures reveal strong enrichment in amino acids such as histidine (H), aspartate/glutamate (D/E), and lysine (K). These residues are well-known in biochemistry for their central roles in catalytic and binding functions. In enzymes like transferases, such residues commonly appear together in conserved regions that drive the chemical reaction. Histidine often acts as a proton donor or acceptor, enabling it to assist in transferring chemical groups during catalysis. Aspartate and glutamate, on the other hand, are acidic residues that help stabilize charged intermediates or bind metal ions that assist in the reaction. Lysine, being positively charged, frequently anchors negatively charged molecules such as phosphate groups or cofactors.  

Several of the motifs identified in our models, such as HIDRL, DEEVK, and HQLLH, closely resemble sequence patterns known to occur in the active or binding sites of transferase enzymes. For example, motifs containing histidine and aspartate together (H–D pairs) are characteristic of catalytic dyads that help in proton transfer reactions, while those containing glutamate and lysine (DEEVK) may reflect regions that coordinate metal ions or cofactors. The presence of these motifs across different architectures strongly suggests that the models are not merely memorizing data, but are recognizing real biochemical patterns associated with enzyme activity.  

The models, however, differ slightly in how they perceive these motifs. The CNN primarily identified multiple short, high-scoring motifs scattered throughout the sequence. This reflects how convolutional filters act as motif detectors, each sensitive to specific short subsequences. The CNN-BiLSTM model captured a balance between these local motifs and their broader sequence context, highlighting not only where the motifs occur but also how they interact along the protein sequence. The BiLSTM alone tended to spread its attention over a wider region, assigning moderate importance to several connected motifs, showing its ability to integrate long-range dependencies. In contrast, the CNN-Attention model concentrated most of its focus on one narrow region (SAEKI), indicating that the attention layer pinpointed a single subsequence as the dominant discriminative feature for transferases.  

Despite these differences in focus, it is notable that all models consistently emphasized certain recurring motifs such as HIDRL, HQLLH, and LLHEL. These common motifs likely correspond to functionally important regions shared among many transferases, possibly aligning with catalytic loops or substrate-binding pockets. The convergence of these findings across independent model architectures strengthens the biological credibility of the learned representations. It implies that, even through different computational mechanisms, each model has independently recognized the same biochemical “signatures” that define the transferase class.  

Overall, the interpretability results suggest that the deep learning models are indeed learning meaningful molecular features rather than arbitrary patterns. The motifs identified are consistent with established biochemical principles—residues responsible for catalysis, cofactor binding, or stabilization of reactive intermediates. In essence, the explainability analysis bridges the gap between model decision-making and biological understanding, showing that the networks capture the same molecular logic that governs enzyme function in nature.

\section{Conclusion}
\label{sec:conclusion}

This study developed a deep learning-based framework for protein functional group classification using sequences from the Protein Data Bank (PDB). Four baseline architectures were implemented, including the Convolutional Neural Network (CNN), Bidirectional Long Short-Term Memory (BiLSTM), CNN-BiLSTM hybrid, and CNN with Attention. These models were used to analyze how different neural architectures capture biochemical information from amino acid sequences. Among them, the CNN achieved the highest validation accuracy of 91.8\%, closely followed by the BiLSTM and the hybrid models. The results indicate that both local motif extraction and sequential dependency modeling contribute significantly to the accurate classification of protein functions.

Explainable AI methods were employed to interpret model predictions and uncover biologically meaningful sequence patterns. Grad-CAM and Integrated Gradients revealed recurring motifs enriched in histidine, aspartate or glutamate, and lysine, which are residues commonly associated with catalytic and metal-binding sites, in enzymes such as transferases. The recurrence of motifs like HIDRL, HQLLH, and DEEVK across multiple architectures confirmed that the networks learned conserved biochemical features rather than arbitrary patterns. Each architecture exhibited distinct interpretive behavior: CNNs emphasized multiple localized motifs, BiLSTMs distributed relevance across broader sequence regions, and attention-based models focused sharply on a single discriminative subsequence.

Overall, the results demonstrate that optimized deep learning architectures can achieve high predictive accuracy while maintaining strong biological interpretability. The integration of explainable AI techniques bridges computational performance with molecular understanding, enabling the identification of motifs consistent with known catalytic mechanisms. Future work will focus on extending this framework to multi-functional enzyme prediction and validating identified motifs against structural and biochemical databases to further strengthen the connection between model explanations and experimentally verified protein function.

\nocite{*}
\bibliographystyle{IEEEtran}

\clearpage

\end{document}